\documentclass[prc,twocolumn,showpacs]{revtex4}
\usepackage{graphicx}
\usepackage{float}

\begin{document}

\title {Spurious Shell Closures in the Relativistic Mean Field Model}

\author{
L. S. Geng$^{1,2,3}$,  J. Meng$^{1}$, H. Toki$^{3}$, W. H.
Long$^{1}$ and G. Shen$^1$}
\affiliation{
$^1$School of Physics, Peking University, Beijing 100871, China \\
$^{2}$Institute of Theoretical Physics, Chinese Academy of
Sciences, Beijing 100080, China\\
 $^3$Research Center for Nuclear
Physics (RCNP), Osaka University, Ibaraki 567-0047, Japan
 }

\begin{abstract}
Following a systematic theoretical study of the ground-state
properties of over 7000 nuclei from the proton drip line to the
neutron drip line in the relativistic mean field model [Prog.
Theor. Phys. 113\,(2005)\,785], which is in fair agreement with
existing experimental data, we observe a few spurious shell
closures, i.e. proton shell closures at $Z=58$ and $Z=92$. These
spurious shell closures are found to persist in all the effective
forces of the relativistic mean field model, e.g. TMA, NL3, PKDD
and DD-ME2.

\end{abstract}

 \maketitle

 The relativistic mean field (RMF) model has achieved great successes
 in studies of nuclear structure for many years.\cite{Walecka74,Reinhard89,Ring96}
 Recently, great efforts in the nuclear physics community have
 been taken to study exotic nuclei with extreme isospin ratios. In
 this respect, the RMF model is considered to be a very promising one due to
 its self-consistent description of the spin-orbit interaction.\cite{Meng05,Vretenar05}
A careful and systematic study of the ground-state
 properties of over 7000 nuclei throughout the periodic table has been performed using
 the RMF+BCS model.\cite{Geng05PTP,Geng05} It was shown that the RMF model
 can describe all the ground-state properties of those experimentally known nuclei
 very well. Nevertheless, a few deficiencies were observed: For example, the neutron skin thickness of $^{208}$Pb
 is predicted to be 0.26 fm; while recent experimental data seem to be favorable to a smaller value. This deficiency
 has been (partially) removed by some lately developed effective forces.\cite{Long04,Lala05,Piekarewicz05}
 Another deficiency, which has not received enough attention, is the appearance of spurious
 shell-closures
 at $Z=58$ and $Z=92$. In Ref. \cite{Geng05PTP}, nuclear binding energies
 around $^{140}$Ce and $^{218}$U were found to be strongly
 overestimated. A closer examination of the corresponding
 single-particle spectra showed that they are caused by spurious
 shell closures at $Z=58$ and $Z=92$ (just beyond the conventional $Z=50$ and $Z=82$
 shell closures), which are further amplified by the co-occurrence
 of the $N=82$ and $N=126$ neutron shell closures.

 In this letter, we investigate whether the same spurious shell closures exist in
 other representative effective forces of the RMF model except TMA \cite{Geng05PTP}. To save space, the formulation of the
 RMF model is not presented here, which can be
 easily found in many review articles (see, for example, Ref.
 \cite{Meng05}). The effective forces we examine in this
 work are TMA,\cite{Sugahara94} which has been demonstrated to work very well throughout
 the periodic table,\cite{Geng05PTP,Geng05} NL3,\cite{Lala97} one of
 most-studied effective forces, PKDD,\cite{Long04} and
 DD-ME2,\cite{Lala05} two lately developed density-dependent effective forces which
 have made some improvements compared to older ones. We
 believe that these four effective forces can represent most
 existing forces in the current RMF
 framework. Since an extensive study such as that performed in Ref.
 \cite{Geng05PTP} is not possible for all the four forces, we would like to concentrate on
 $^{140}$Ce and $^{218}$U,
 where the spurious shell closures at $Z=58$ and 92 were first clearly identified (see Fig.\,2 of Ref. \cite{Geng05PTP}). For comparison, we also study
 $^{132}$Sn and $^{208}$Pb, which are used as reference nuclei. To
 avoid any possible ambiguity, all the results shown in this study are
 obtained with a spherical code and without the pairing correlation.
 This first approximation is due to the fact that these nuclei
 are found to be spherical,\cite{Geng05PTP,Lala99} and the second
 approximation (together with the first) implies that the so obtained binding energies for these nuclei are at
 the lower bound.

 \begin{table}[b]
\setlength{\tabcolsep}{0.6 em} \caption{Binding energies of
$^{132}$Sn, $^{208}$Pb, $^{140}$Ce and $^{218}$U obtained with
effective forces TMA, NL3, PKDD and DD-ME2 (in units of MeV). The
experimental data are taken from Ref. \cite{Audi03}. }
\begin{center}\label{table1}
\begin{tabular}{c@{\hspace{2ex}}|cccc}
\hline\hline
 &$^{132}$Sn&$^{208}$Pb&$^{140}$Ce&$^{218}$U\\
\hline
TMA&1103.4&1634.7&1179.4&1674.0\\
NL3&1103.2&1637.5&1176.3&1674.7\\
PK-DD&1102.6&1637.2&1179.0&1677.7\\
DD-ME2&1103.5&1639.0&1175.9&1673.5\\
\hline
Exp.&1102.9&1636.4&1172.7&1665.6\\
 \hline\hline
\end{tabular}
\end{center}
\end{table}
 \begin{table*}[t]
\setlength{\tabcolsep}{0.6 em} \caption{Proton shell closures at
$Z=50$, 58 in $^{132}$Sn ($^{140}$Ce) and those at $Z=82$, 92 in
$^{208}$Pb ($^{218}$U) obtained with effective forces TMA, NL3,
PKDD and DD-ME2 (in units of MeV). The first column under
$^{132}$Sn ($^{140}$Ce) is the shell closure at $Z=50$ and the
second column is that at $Z=58$. The first column under $^{208}$Pb
($^{218}$U) is the shell closure at $Z=82$ while the second column
is that at $Z=92$.}
\begin{center}\label{table2}
\begin{tabular}{c@{\hspace{2ex}}|cc|cc|cc|cc}
\hline\hline
 &\multicolumn{2}{c|}{$^{132}$Sn}&\multicolumn{2}{c|}{$^{140}$Ce}&\multicolumn{2}{c|}{$^{208}$Pb}&\multicolumn{2}{c}{$^{218}$U}\\
\hline
TMA&5.66&3.73&5.36&4.12&2.47&4.01&2.34&3.93\\
NL3&6.15&2.81&5.78&3.50&3.50&3.20&3.14&3.54\\
PKDD&6.45&2.99&6.09&3.63&3.59&3.36&3.25&3.63\\
DD-ME2&6.45&2.39&6.01&3.08&4.06&2.74&3.64&3.28\\
\hline
 Exp.&6.03&0.96&&&4.20&0.90\\
 \hline\hline
\end{tabular}
\end{center}
\end{table*}
 In Table \ref{table1}, the binding energies of $^{132}$Sn,
 $^{140}$Ce, $^{208}$Pb and $^{218}$U obtained using the effective
 forces TMA, NL3, PKDD and DD-ME2 are tabulated. It is easily seen
 that all the four forces overestimate the binding energy of $^{218}$U by at least 8\,MeV;
 while only TMA and PKDD overestimate the binding energy of $^{140}$Ce by the same
 order. This is in agreement with the systematic study of Ref.
 \cite{Geng05PTP}. Among the four forces, the latest force DD-ME2 provides the best
 description of the experimental data, but it still fails for
 $^{218}$U. This corresponds to the spurious shell closure at $Z=92$,
 as we shall see in the following. It is interesting to
 note that PKDD does particularly well for $^{132}$Sn and $^{208}$Pb while it
 is not so good for $^{140}$Ce and $^{218}$U. Considering that these effective forces
 describe most nuclei throughout the periodic table with a
 discrepancy of 2--3\,MeV,\cite{Geng05PTP,Lala99} the 8\,MeV
 discrepancy is rather large. This therefore indicates some
 physics that should not be ignored. Following Bertsch \textit{et
 al},\cite{Bertsch05} these nuclei should be thought of as ``critical
 nuclei'' of the corresponding forces.

 In Ref. \cite{Geng05PTP}, the overbindings of
 $^{140}$Ce and $^{218}$U were attributed to the corresponding spurious shell closures
 at $Z=58$ and $Z=92$. In the
 following, we investigate whether it is still the case for NL3,
 PKDD and DD-ME2. The proton shell closures at $Z=50$, 58 in $^{132}$Sn
 ($^{140}$Ce) and those at
 $Z=82$, 92 in  $^{208}$Pb ($^{218}$U)
 obtained with effective forces TMA, NL3, PKDD and DD-ME2 are tabulated in
 Table \ref{table2}. For comparison, the
 experimental data for $^{132}$Sn and $^{208}$Pb are also shown.
 For $^{132}$Sn, it is seen that NL3 agrees with experiment best; TMA
 underestimates the $Z=50$ shell closure but overestimates the
 $Z=58$ shell closure; while PKDD and DD-ME2 are of similar
 quality. The large $Z=58$ shell closure in TMA and PKDD persists even at $^{140}$Ce,
 which explains the overbinding of this nucleus as listed in Table \ref{table1}.
  For $^{208}$Pb, DD-ME2 agrees with experiment best; TMA
 underestimates the $Z=82$ shell closure but overestimates
 the $Z=92$ shell closure; while NL3 and PKDD are of similar
 quality. For $^{218}$U, all the effective forces predict a very
 large shell closure at $Z=92$: it is even larger than that at
 $82$ in TMA, NL3 and PKDD; while in DD-ME2 it is slightly smaller
 than that at $82$. This extremely large $Z=92$ shell closure is
 believed to be responsible for the overbinding of $^{218}$U as listed in Table \ref{table1}.
 It is interesting to note that the existence of these spurious shell
 closures seems to be a common feature of the current RMF forces (at least for those investigated here), while in
 nonrelativistic Hartree-Fock models, no such a common feature has been found.

The nucleus is composed of both protons and neutrons. Therefore,
one must study not only protons but also neutrons to obtain a
complete picture. In Table \ref{table3}, the neutron shell
closures at $N=82$ in $^{132}$Sn ($^{140}$Ce) and those at $N=126$
in $^{208}$Pb ($^{218}$U) obtained with effective forces TMA, NL3,
PKDD and DD-ME2 are tabulated. It is seen that all the effective
forces overestimate the $N=82$ shell closure by about 1.5\,MeV.
From $^{132}$Sn to $^{140}$Ce, this shell closure increases in
NL3, PKDD and DD-ME2; while it decreases in TMA. For $^{208}$Pb,
TMA agrees with
 experiment very well; while all the others overestimate the
 experimental value with DD-ME2 having the largest discrepancy. From $^{208}$Pb to $^{218}$U,
 a decrease of about 1.5\,MeV is seen for all the effective forces
 . From Table \ref{table2} and
 Table \ref{table3}, it can be concluded that the better description of
 the proton shell closure of $^{208}$Pb in NL3, PKDD and DD-ME2 compared to TMA is at the
 expense of the overestimation of the neutron shell closure, which is particularly true for DD-ME2.

\begin{table}[t]
\setlength{\tabcolsep}{0.6 em} \caption{Neutron shell closures at
$N=82$ in $^{132}$Sn ($^{140}$Ce) and those at $N=126$ in
$^{208}$Pb ($^{218}$U) obtained with effective forces TMA, NL3,
PKDD and DD-ME2 (in units of MeV).}
\begin{center}\label{table3}
\begin{tabular}{c@{\hspace{2ex}}|cccc}
\hline\hline
 &$^{132}$Sn&$^{140}$Ce&$^{208}$Pb&$^{218}$U\\
\hline
TMA&6.78&5.58&3.58&2.14\\
NL3&6.17&6.38&4.60&3.07\\
PKDD&6.50&6.65&4.78&3.16\\
DD-ME2&6.17&7.07&5.25&3.72\\
\hline
 Exp.&4.84&&3.43&\\
 \hline\hline
\end{tabular}
\end{center}
\end{table}

The importance of identifying these spurious shell closures or
``critical nuclei" can be summarized as follows: (i) It points out
the regions in the periodic table that the present RMF model can
not describe so well. (ii) It tells us to further study what
causes these spurious shell closures and how to remove them. As a
result, the current formulation of the RMF model can be improved.
In the following, we discuss several approaches that might be
helpful to remove the observed spurious shell closures. It should
be noted that our discussions are restricted to the mean-field
level and are not meant to be exhaustive. First, one may need a
more balanced fitting strategy. For example, only the ground-state
properties of twelve nuclei are fitted to obtain the latest
effective force DD-ME2. Meanwhile these twelve nuclei are in some
sense neutron rich. Since all the current effective forces in the
RMF model adopt a similar procedure to fix their parameter values,
it is unsurprising that they all overestimate $^{218}$U, which is
proton rich. Although an extensive fitting strategy such as that
used in obtaining the latest series of Hartree-Fock-Bogoliubov
mass tables
 \cite{Samyn04} in the RMF model is not likely to be carried out in the near future, a more balanced
fitting strategy can hopefully remove or at least reduce the
overbinding of $^{218}$U and those around it. In this sense,
taking into account the single-particle spectra, in particular
those of $^{208}$Pb, during the fitting process might also improve
the description of these critical nuclei.

Another alternative is to make magic nuclei slightly less bound.
The reason is that particle-number conserved pairing methods
usually give a few MeV of energy gain to even magic
nuclei.\cite{Geng05PRC} Therefore, fitting the binding energies of
magic nuclei exactly to the experimental data or even slightly
larger than the experimental data (most current RMF forces behave
in this way) leaves no room for further improvement and meanwhile
overestimates magic number effects. Just to see how we can reduce
the overbindings of $^{140}$Ce and $^{218}$U by making magic
nuclei, particularly $^{132}$Sn and $^{208}$Pb, less bound, we
modify the effective force DD-ME2. It should be noted that the
purpose is not to develop a new effective force, but to
demonstrate our above statements. To keep things simple, we only
change the nucleon masses. The nucleon masses used in DD-ME2 are
939.0\,MeV for both protons and neutrons.\cite{Ring05} We reduce
them to 938.5\,MeV. The binding energies obtained with this
modified DD-ME2 (denoted as DD-ME2(M)) for $^{132}$Sn, $^{140}$Ce,
$^{208}$Pb and $^{218}$U are tabulated in Table \ref{table4}. One
can easily see that now the discrepancies are more evenly
distributed, i.e. serious discrepancies disappear.

 \begin{table}[t]
\setlength{\tabcolsep}{0.6 em} \caption{Binding energies of
$^{132}$Sn, $^{208}$Pb, $^{140}$Ce and $^{218}$U obtained with
effective forces DD-ME2 and DD-ME2(M) (in units of MeV). The
experimental data are taken from Ref. \cite{Audi03}.}
\begin{center}\label{table4}
\begin{tabular}{c@{\hspace{2ex}}|cccc}
\hline\hline
 &$^{132}$Sn&$^{208}$Pb&$^{140}$Ce&$^{218}$U\\
\hline
DD-ME2&1103.5&1639.0&1175.9&1673.5\\
 DD-ME2(M)&1100.6&1634.4&1172.9&1168.7\\
\hline
Exp.&1102.9&1636.4&1172.7&1665.6\\
 \hline\hline
\end{tabular}
\end{center}
\end{table}

To make RMF calculations more reliable for astrophysical studies
and/or studies of exotic nuclei, the predictions of nuclear masses
must be improved. It may or may not be possible within the current
formulation of the RMF model. One may finally resort to a similar
procedure adopted in non-relativistic calculations.\cite{Samyn04}
Although recently constructed effective forces do show
improvements compared to older ones, deficiencies remain as
demonstrated in this work. The recent systematic study
\cite{Geng05PTP} showed that most discrepancies between the RMF
results and existing experimental data actually originate from the
regions surrounding the so-called critical nuclei. Therefore,
removing them should be viewed as an essential criteria for the
next-generation effective forces. These works are underway.

This work is partially supported by the National Natural Science
Foundation of China under Grant Nos. 10435010, 10221003, and
10505002.

\end{document}